\documentclass[journal]{IEEEtran}
\ifCLASSINFOpdf
\else
\fi
\usepackage{amsmath}
\usepackage{graphicx}
\usepackage{color}

\begin{document}
%
\title{A Distributed Algorithm for Solving Linear Algebraic Equations Over Random Networks }
%
%
%

\author{Seyyed~Shaho~Alaviani,~\IEEEmembership{Student Member,}
        and~Nicola~Elia,~\IEEEmembership{Senior Member,~IEEE}
\thanks{Seyyed Shaho Alaviani is with the Department
of Electrical and Computer Engineering, Iowa State University, Ames,
IA, 50011 USA e-mail: shaho@iastate.edu.}
\thanks{Nicola Elia is with the Department
	of Electrical and Computer Engineering, University of Minnesota, Minneapolis,
	MN, 55455 USA e-mail: nelia@umn.edu.}
\thanks{A preliminary version of this paper has appeared without proofs in \cite{alavianiCDC}. This work has been done while Nicola Elia was at Iowa State University.}
\thanks{This work was supported by National Science Foundation under Grant CCF-1320643, Grant CNS-1239319, and AFOSR Grant FA 9550-15-1-0119.}
}

%
%

\markboth{}%
{Alaviani \MakeLowercase{\textit{et al.}}: Dist. Alg. Lin. Algeb. Eq. Rand. Net.}
%



\maketitle

\begin{abstract}
In this paper, we consider the problem of solving linear algebraic equations of the form $Ax=b$ among multi agents which seek a solution by using local information in presence of random communication topologies. The equation is solved by $m$ agents where each agent only knows a subset of rows of the partitioned matrix $[A,b]$. We formulate the problem such that this formulation does not need the distribution of random interconnection graphs. Therefore, this framework includes asynchronous updates or unreliable communication protocols without B-connectivity assumption. We apply the random Krasnoselskii-Mann iterative algorithm which converges almost surely and in mean square to a solution of the problem for any matrices $A$ and $b$ and any initial conditions of agents' states. We demonestrate that the limit point to which the agents' states converge is determined by the unique solution of a convex optimization problem regardless of the distribution of random communication graphs. Eventually, we show by two numerical examples that the rate of convergence of the algorithm cannot be guaranteed. 
\end{abstract}


%
\IEEEpeerreviewmaketitle

\section{Introduction}
Linear algebraic equations arise in modeling of many natural phenomena such as forecasting and estimation \cite{forc}. Since the processors are physically separated from each others, distributed computations to solve linear algebraic equations are important and useful. The linear algebraic equation considered in this paper is of the form $Ax=b$ that is solved simultaneously by $m$ agents assumed to know only a subset of the rows of the partitioned matrix $[ A, b ]$, by using local information from their neighbors; indeed, each agent only knows $A_{i}x_{i}=b_{i}, i=1,2,...,m$, where the goal of them is to achieve a consensus $x_{1}=x_{2}=...=x_{m}=\tilde{x}$ where $\tilde{x} \in \{ \bar{x} | \bar{x}=\underset{x}{\arg\min} \Vert Ax-b \Vert \}$. Several authors have  proposed algorithms for solving the problem over non-random networks \cite{lae1}-\cite{lae14}. This problem can also be solved by subgradient algorithm proposed in \cite{gradie1}. Other distributed algorithms for solving linear algebraic equations have been proposed by some investigators \cite{elia}-\cite{cano} that the problems they consider are not the same as the problem considered in this paper. Some approaches propose cooperative solution methods that exploit the matrix $A$ interconnectivity and have each node in charge of one single solution variable or a dual variable \cite{elia}-\cite{elia1}.   

In practice, because of packet drops or links' failures, random graphs are suitable models for the underlying graph over which agents communicate. Therefore, solving linear algebraic equations over random networks is very important and useful. As mentioned above, each agent $i$ wishes to solve $Ax=b$ by using its own private equation $A_{i}x_{i}=b_{i}$ in presence of random communication with its neighbors. One view of the problem is to formulate it as a constrained consensus problem over random networks and use the result in \cite{shijoh}; nevertheless, the result in \cite{shijoh} needs each agent to use projection onto its constraint set with some probability at each time and also needs weighted matrix of the graph to be independent at each time. Another view of the problem is to formulate it as a distributed convex optimization problem over random networks and use the results in \cite{14}-\cite{dualaver}. Nevertheless, the results in \cite{14}-\cite{dualaver} are based on subgradient descent or diminishing step size that have slow convergence as an optimal solution is approached. Furthermore, the results in \cite{14}-\cite{dualaver} need weighted matrix of the graph to be independent and identically distributed (i.i.d.).

In a \textit{synchronous} protocol, all nodes activate at the same time and perform communication updates. This protocol requires a common notion of time among the nodes. On the other hand, in \textit{asynchronous} protocol, each node has its own concept of time defined by a local timer which randomly triggers either by the local timer or by a message from neighboring nodes. As the dimension of the network increases, synchronization becomes an issue. The above problem can be solved by asynchronous subgradient algorithm proposed in \cite{nedicasy}-\cite{athans}, to cite a few; nevertheless, the authors assume that there exists a bounded time interval such that each edge transmites a message at least once(B connectivity), or nodes/edges activation are i.i.d. Recently, the authors of \cite{laww1} have proposed asynchronous algorithms for solving the linear algebraic equation over time-varying networks where they impose B-connectivity assumption.  

\textbf{Contribution:} In this paper, we consider the problem of solving linear algebraic equations of the form $Ax=b$ over a network of $m$ agents where each agent only knows a subset of the rows of the partitioned matrix $[A,b]$ in presence of random communication graphs. Several authors in the literature have considered solving linear algebraic equations over switching networks with B-connectivity assumption such as \cite{laww1}. However, B-connectivity assumption is not guaranteed to be satisfied for random networks. We formulate this problem such that this formulation does not need the distribution of random communication graphs or B-connectivity assumption if the weighted matrix of the graph is doubly stochastic. Thus this formulation includes asynchronous updates or unreliable communication protocols. We assume that the set $\mathcal{S}=\{ x | \displaystyle  \min_{x} \Vert Ax-b \Vert=0  \}$ is nonempty. Since the Picard iterative algorithm may not converge, we apply the random Krasnoselskii-Mann iterative algorithm for converging almost surely and in mean square\footnote{Although this paper is the completed version of \cite{alavianiCDC}, we prove here that the algorithm also converges in mean square to the solution.} to a point in $\mathcal{S}$ for \textit{any} matrices $A$ and $b$ and \textit{any} initial conditions. The proposed algorithm, like those of \cite{lae1}-\cite{laddd} and \cite{laww2}-\cite{lae13}, requires that whole solution vector is computed and exchanged by each node over a network. Based on initial conditions of agents' states, we show that the limit point to which the agents' states converge is determined by the unique solution of a feasible convex optimization problem independent from the distribution of random links' failures. 

The paper is organized as follows. In section II, some preliminaries are given. In section III, formulations of the problem are presented. In section IV, the main results of this paper are presented. Finally, two numerical examples are given to show that the rate of convergence of the algorithm cannot be guaranteed.  

\textit{Notations:} $\Re$ denotes the set of all real  numbers. We use 2-norm for vectors and induced 2-norm for matrices, i.e., for any vector $z \in \Re^{n},\Vert z \Vert=\Vert z \Vert_{2}=\sqrt{z^{T}z},$ and for any matrix $Z \in \Re^{n \times n},\Vert Z \Vert=\Vert Z \Vert_{2}=\sqrt{\lambda_{max}(Z^{T}Z)}=\sigma_{max}(Z)$ where $T$ represents the transpose of matrix $Z$, $\lambda_{max}$ represents maximum eigenvalue, and $\sigma_{max}$ represents largest singular value. For any matrix $Z \in \Re^{n \times n}$ with $Z=[z_{ij}]$,  $\Vert Z \Vert_{1}= max_{1 \leq j \leq n} \{\sum_{i=1}^{n} z_{ij} \}$ and $\Vert Z \Vert_{\infty}= max_{1 \leq i \leq n} \{\sum_{j=1}^{n} z_{ij} \}$. Sorted in an increasing order, $\lambda_{2}(Z)$ represents the second eigenvalue of a matrix $Z$. $Re(r)$ represents the real part of the complex number $r$. $I_{n}$ represents Identity matrix of size $n \times n$ for some $n \in N$ where $N$ denotes the set of all natural numbers. $\otimes$  denotes the Kronecker product. $\emptyset$ represents the empty set. $\textbf{0}_{n}$ represents the vector of dimension $n$ whose entries are all zero. $\textbf{1}_{n}$ represents the vector of dimension $n$ whose entries are all one. $E[x]$ denotes Expectation of random variable $x$.

\section{Preliminaries}

A vector $v \in \Re^{n}$ is said to be a \textit{stochastic vector} when its components $v_{i}, i=1,2,...,n$, are non-negative and their sum is equal to 1; a square $n \times n$ matrix $V$ is said to be a \textit{stochastic matrix} when each row of $V$ is a stochastic vector. A square $n \times n$ matrix $V$ is said to be \textit{doubly stochastic} matrix when both $V$ and $V^{T}$ are stochastic matrices.

Let $X$ be a real Hilbert space with norm $\Vert . \Vert $ and inner product $<.,.>$. Let $C$ be a nonempty subset of the Hilbert space $X$ and $H:C \longrightarrow X$. The point $\bar{x}$ is called a \textit{fixed point} of $H$ if $\bar{x}=H(\bar{x})$. The set of fixed points of $H$ is represented by $Fix(H)$. 

Let $(\Omega^{*},\sigma)$ be a measurable space ($\sigma$-sigma algebra) and $C$ be a nonempty subset of a Hilbert space $X$. A mapping $x:\Omega^{*} \longrightarrow X$ is \textit{measurable} if $x^{-1}(U) \in \sigma$ for each open subset $U$ of $X$. The mapping $T:\Omega^{*} \times C \longrightarrow X$ is a \textit{random map} if for each fixed $z \in C$, the mapping $T(.,z):\Omega^{*} \longrightarrow X$ is measurable, and it is \textit{continuous} if for each $\omega^{*} \in \Omega^{*}$ the mapping $T(\omega^{*},.):C \longrightarrow X$ is continuous.

Let $\omega^{*}$ and $\omega$ denote elements in the sets $\Omega^{*}$ and $\Omega$, repectively.

\textbf{Definition 1} \cite{alavianiTAC}-\cite{alaviani}: A point $\hat{x} \in X$  is a \textit{fixed value point} of a random map $T$ if $\hat{x}=T(\omega^{*},\hat{x})$ for all $\omega^{*} \in \Omega^{*}$, and $FVP(T)$ represents the set of all fixed value points of $T$. 

\textbf{Definition 2:} Let $C$ be a nonempty subset of a Hilbert space $X$ and $T:\Omega^{*} \times C \longrightarrow C$ be a random map. The map $T$ is said to be \textit{non-expansive random operator} if for each $\omega^{*} \in \Omega^{*}$ and for arbitrary $x,y \in C$ we have 
$$\Vert T(\omega^{*} ,x)-T(\omega^{*} ,y) \Vert \leq \Vert x-y \Vert.$$

\textbf{Definition 3:} Let $C$ be a nonempty subset of a Hilbert space $X$ and $H:C \longrightarrow C$ be a map. The map $H$ is said to be \textit{non-expansive} if for arbitrary $x,y \in C$ we have 
$$\Vert H(x)-H(y) \Vert \leq \Vert x-y \Vert.$$

\textbf{Remark 1} \cite{alavianiTAC}-\cite{alaviani}: Let $C$ be a closed convex subset of a Hilbert space $X$. The set of fixed value points of a non-expansive random operator $T:\Omega^{*} \times C \longrightarrow C$ is closed and convex.

\textbf{Definition 4:} A sequence of random variables $x_{n}$ is said to \textit{converge almost surely} to $x$ if there exists a set $A$ such that $Pr(A)=0$, and for every $\omega \notin A$
$$lim_{n \longrightarrow \infty} \Vert x_{n}(\omega)-x(\omega) \Vert=0.$$

\textbf{Definition 5:} A sequence of random variables $x_{n}$ is said to \textit{converge in mean square} to $x$ if 
$$E [\Vert x_{n}-x \Vert^{2}] \longrightarrow 0 \quad \text{as $n \longrightarrow \infty$}.$$

\textbf{Lemma 1} \cite{threedec}: Let $W \in \Re^{m \times m}$. Then $\Vert W \Vert_{2} \leq \sqrt{\Vert W \Vert_{1} \Vert W \Vert_{\infty}}$.

\textbf{Definition 6} \cite{fejer0}: Suppose $C$ is a closed convex nonempty set and $\{ x_{n} \}_{n=0}^{\infty}$ is a sequence in $X$. $\{ x_{n} \}_{n=0}^{\infty}$ is said to be \textit{Fej\'{e}r monotone with respect to $C$} if

$$\Vert x_{n+1}-z \Vert \leq \Vert x_{n}-z\Vert, \quad{} \forall z \in C, n \geq 0.$$

\textbf{Lemma 2} \cite{fejer0}: Suppose the sequence $\{ x_{n} \}_{n=0}^{\infty}$ is Fej\'{e}r monotone with respect to $C$. Then $\{ x_{n} \}_{n=0}^{\infty}$ is bounded.

\textbf{Lemma 3} \cite{fejerbook}: Let $\{ x_{n} \}_{n=0}^{\infty}$ be a sequence in $X$ and let $C$ be a closed affine subspace of $X$. Suppose that $\{ x_{n} \}_{n=0}^{\infty}$ is Fej\'{e}r monotone with respect to $C$. Then $P_{C} x_{n}=P_{C} x_{0}, \forall n \in N,$ where $P_{C}$ denote projection onto the set $C$.

\textbf{Remark 2} \cite[Ch. 2]{funct}: Due to strict convexity of the norm $\Vert . \Vert$ in a Hilbert space $X$, if $\Vert x \Vert = \Vert y \Vert = \Vert (1-\beta) x + \beta y \Vert$ where $x,y \in X$ and $\beta \in (0,1),$ then $x=y$. 

\textbf{Proposition 1} \cite{royden}: $\Re$ is a closed set.

\textbf{Definition 7} \cite{boyed}: A set $C \subseteq \Re^{n}$ is \textit{affine} if the line through any two distinct points in $C$ lies in $C$, i.e., if for any $z,y \in C$ and $\alpha \in \Re$, we have $\alpha z +(1-\alpha) y \in C$. 

\textbf{Remark 3} \cite{boyed}: If $C$ is an affine set and $z_{0} \in C$, then the set
$$C-z_{0}=\{ z-z_{0} |z \in C \}$$
is a subspace.

\textbf{Definition 8:} The map $T:X \longrightarrow X$ is said to be \textit{firmly nonexpansive} if for each $x,y \in X$,
$$\Vert T(x)-T(y) \Vert^{2} \leq <T(x)-T(y),x-y>.$$

\textbf{Remark 4} \cite{image}: $\phi:X \longrightarrow X$ is a firmly nonexpansive mapping if $T:X \longrightarrow X$ is a nonexpansive mapping where
$$\phi(x)=\frac{1}{2} (x+T(x)).$$
Moreover, every firmly nonexpansive mapping is nonexpansive by Cauchy--Schwarz inequality.

\textbf{Lemma 4} \cite{finitehilbert}: Let $\phi_{i}:X \longrightarrow X, i=1,2,. . . ,\tilde{N}$, be firmly nonexpansive with $\cap_{i=1}^{\tilde{N}} Fix(\phi_{i}) \neq \emptyset$, where $X$ is finite dimensional. Then the random sequence generated by
\begin{equation}\label{projalg}
x_{0} \in D \quad{\text{arbitrary}}, \quad{} x_{n+1}=\phi_{r(n)}(x_{n}), n \geq 0,
\end{equation}
where each element of $\{ 1, . . ., \tilde{N}\}$ appears in the sequence $\{ r(0),r(1),. . . \}$ an infinite number of times, converges to some point in $\cap_{i=1}^{\tilde{N}} Fix(\phi_{i})$.

\textbf{Lemma 5} \cite{royden} \textit{(Fatou's Lemma)}: If $\tau_{n}:\Omega \longrightarrow [0, \infty]$ is measurable, for each positive integer $n$, then
$$\int_{\Omega} (\liminf_{n \longrightarrow \infty} \tau_{n}) d\mu \leq \liminf_{n \longrightarrow \infty} \int_{\Omega} \tau_{n} d\mu.$$  

\textbf{Lemma 6} \cite{royden} \textit{(The Lebesque Dominated Convergence Theorem)}: Let $\{ \tau_{n} \}$ be a sequence of measurable functions on $\Omega$. Suppose there is a function $g$ that is integrable over $\Omega$ and dominates $\{ \tau_{n} \}$ on $\Omega$ in the sense that $\vert \tau_{n} \vert \leq g$ on $\Omega$ for all $n$. If $\{ \tau_{n} \} \longrightarrow \tau$ almost surely on $\Omega$, then $\tau$ is integrable over $\Omega$ and $\displaystyle \lim_{n \longrightarrow \infty} \int_{\Omega} \tau_{n}=\int_{\Omega} \tau$.

\section{Problem Formulation}

Now, we define the problem, considered in this paper, of solving linear algebraic equations over a random network. We adopt the following paragraph from \cite{alavianiTAC}.

A network of $m$ nodes labeled by the set $\mathcal{V}=\lbrace 1,2,...,m \rbrace $ is considered. The topology of the interconnections among nodes is not fixed but defined by  a set of graphs $\mathcal{G}(\omega^{*})=(\mathcal{V},\mathcal{E}(\omega^{*}))$ where $\mathcal{E}(\omega^{*})$ is the ordered edge set $\mathcal{E}(\omega^{*}) \subseteq \mathcal{V} \times \mathcal{V}$ and $\omega^{*} \in \Omega^{*}$ where $\Omega^{*}$ is the set of all possible communication graphs, i.e., $\Omega^{*}=\lbrace \mathcal{G}_{1}, \mathcal{G}_{2}, ..., \mathcal{G}_{\bar{N}}  \rbrace$. We assume that $(\Omega^{*},\sigma)$ is a measurable space where $\sigma$ is the $\sigma$-algebra on $\Omega^{*}$. We write $\mathcal{N}_{i}^{in} (\omega^{*})/\mathcal{N}_{i}^{out} (\omega^{*})$ for the labels of agent $i$'s in/out neighbors at graph $\mathcal{G}(\omega^{*})$ so that there is an arc in $\mathcal{G}(\omega^{*})$ from vertex $j/i$ to vertex $i/j$ only if agent $i$ receives/sends information from/to agent $j$. We write $\mathcal{N}_{i}(\omega^{*})$ when $\mathcal{N}_{i}^{in}(\omega^{*})=\mathcal{N}_{i}^{out}(\omega^{*})$. We assume that there are no self-looped arcs in the communication graphs.

The agents want to solve the problem $\displaystyle  \min_{x} \Vert Ax-b \Vert, A \in \Re^{\mu \times q}, b \in \Re^{\mu}$, where each agent merely knows a subset of the rows of the partitioned matrix $[A,b]$; precisely, each agent knows a private equation $A_{i}x_{i}=b_{i}, i=1,2,...,m,$ where $A_{i} \in \Re^{\mu_{i} \times q}, b_{i} \in \Re^{\mu_{i}}, \sum_{i=1}^{m} \mu_{i}=\mu$. We also assume that there is no communication delay or noise in delivering a message from agent $j$ to agent $i$.

Similar to \cite{alavianiTAC}, we define the weighted graph matrix $\mathcal{W}(\omega^{*})=[\mathcal{W}_{ij}(\omega^{*})]$  as $\mathcal{W}_{ij}(\omega^{*})=a_{ij}(\omega^{*})$ for $j \in \mathcal{N}_{i}^{in}(\omega^{*}) \cup \{ i \}$ and $\mathcal{W}_{ij}(\omega^{*})=0$ otherwise, where $a_{ij}(\omega^{*})>0$ is the scalar constant weight that agent $i$ assigns to the information $x_{j}$ received from agent $j$. For instance, if $\mathcal{W}(\mathcal{G}_{k})=I_{m}$, for some $1 \leq k \leq \bar{N}$, implies that there is no edges in $\mathcal{G}_{k}$ or/and all nodes are not activated for communication updates in asynchronous protocol.

Now we impose the following assumption on the weights.

\textbf{Assumption 1} \cite{alavianiTAC}: The weighted graph matrix $\mathcal{W}(\omega^{*})$ is doubly stochastic for each $\omega^{*} \in \Omega^{*}$, i.e.,

\textit{i)} $\sum_{j \in \mathcal{N}_{i}^{in}(\omega^{*}) \cup \{ i \}} \mathcal{W}_{ij}(\omega^{*})=1, i=1,2,...,m$,

\textit{ii)} $\sum_{j \in \mathcal{N}_{i}^{out}(\omega^{*}) \cup \{ i \}} \mathcal{W}_{ij}(\omega^{*})=1, i=1,2,...,m.$

\textbf{Remark 5} \cite{alavianiTAC}: Although double-stochasticity is restrictive in distributed setting \cite{restrictive}, it is shown in \cite{alavianiTAC} that Assumption 1 allows us to remove the distribution of random interconnection graphs. Moreover, any networks with undirected links satisfies Assumption 1.

The objective of each agent is to collaboratively seek the solution of the following optimization problem using local information:

$$\min \sum_{i=1}^{m} \Vert A_{i}x-b_{i} \Vert^{2}$$
where $x \in \Re^{q}$. Now we impose the following assumption on the underlying graph.

\textbf{Assumption 2} \cite{alavianiTAC}: The union of all of the graphs in $\Omega^{*}$ is strongly connected, i.e., $Re[\lambda_{2}(\sum_{\omega^{*} \in \Omega^{*}} (I_{m}-\mathcal{W}(\omega^{*})))]>0$. 

\textbf{Remark 6:} The statement of Assumption 2 is a necessary and sufficient condition (see \cite{alavianiTAC}).

Assumption 2 ensures that the information sent from each node will be finally obtained by every other node through a directed path.  Next we formulate the above problem as follows.

\textbf{Problem 1:} Let $T(\omega^{*},x):=W(\omega^{*})x, \omega^{*} \in \Omega^{*}$ where $W(\omega^{*}):=\mathcal{W}(\omega^{*}) \otimes I_{q}, \omega^{*} \in \Omega^{*}$. Then the above problem under Assumptions 1 and 2 can be formulated as follows:
\begin{equation}\label{1}
\begin{aligned}
& \underset{x}{\text{min}}
& & f(x)=\sum_{i=1}^{m} \Vert A_{i}x_{i}-b_{i} \Vert^{2} \\
& \text{subject to}
& & x \in FVP(T),
\end{aligned}
\end{equation}
where $x=[x_{1},...,x_{m}]^{T}, x_{i} \in \Re^{q}, i=1,2,...,m.$

\textbf{Definition 9} \cite{alavianiTAC}: Given a weighted graph matrix $\mathcal{W}(\omega^{*})$,  $T(\omega^{*},x):=W(\omega^{*})x$, $\forall \omega^{*} \in \Omega^{*},$ is said to be \textit{weighted random operator of the graph}. Similarly, for non-random case, $T(x):=Wx$ is said to be \textit{weighted operator of the graph}. 

\textbf{Remark 7:} The set $\mathcal{C}=\{ x \in \Re^{mq} | x_{i}=x_{j}, 1 \leq i,j \leq m, x_{i} \in \Re^{q} \}$ is known as \textit{consensus subspace}. Consensus subspace is in fact the fixed value points set of weighted random operator of the graph with Assumption 2 \cite{alavianiTAC}.

 From Assumption 1 and Lemma 1, the weighted random operator of the graph is nonexpansive \cite{alavianiTAC}. Therefore, according to \cite{alavianiTAC}, the distribution of random communication graphs is not needed. Furthermore, $FVP(T)$ is a convex set (see Remark 1), and Problem 1 is a convex optimization problem. Note that the optimization problem (\ref{1}) is a special case of the proposed optimization problem in \cite{alavianiTAC}-\cite{alaviani}. We mention that the Hilbert space considered in this paper is $(\Re^{mq}, \Vert . \Vert_{2})$

\section{Main Results}

Before presenting our main results, we impose the following assumption on the equation $Ax=b$.

\textbf{Assumption 3:} The linear algebraic equation $Ax=b$ has a solution, namely $\mathcal{S} \neq \emptyset$.

Problem 1 with Assumption 3 can be reformulated as finding $x=[x_{1},x_{2}, ...,x_{m}]^{T}$ such that

\begin{equation}\label{2}
\bar{A}x=\bar{b},
\end{equation}
and
\begin{equation}\label{3}
x \in FVP(T),
\end{equation}
where 
\[\bar{A}=\left(
\begin{array}{c c c c}
A_{1} & 0 & \cdots & 0\\
0 & A_{2} & \cdots & 0\\
\vdots & \cdots & \ddots & \vdots\\
0 & 0 & \cdots & A_{m}\\ 
\end{array}
\right), \bar{b}=\left(
\begin{array}{c c c c}
b_{1}\\
b_{2}\\
\vdots\\
b_{m}\\ 
\end{array}
\right).\]

\textbf{Lemma 7:} The solution set of (\ref{2}) is equal to the solution set of the following equation:  
\begin{equation}\label{4}
\tilde{A} x + \tilde{b}=x,
\end{equation}
where
\begin{equation}\label{5}
\tilde{A}=
\end{equation}
\[\left(
\begin{array}{c c c c}
I_{q}-\theta_{1} A_{1}^{T} A_{1} & 0 & \cdots & 0\\
0 & I_{q}-\theta_{2} A_{2}^{T} A_{2} & \cdots & 0\\
\vdots & \cdots & \ddots & \vdots\\
0 & 0 & \cdots & I_{q}-\theta_{m} A_{m}^{T} A_{m}\\ 
\end{array}
\right),\] 
\begin{equation}\label{6}
\tilde{b}=\left(
\begin{array}{c c c c}
\theta_{1} A_{1}^{T} b_{1}\\
\theta_{2} A_{2}^{T} b_{2}\\
\vdots\\
\theta_{m} A_{m}^{T} b_{m}\\ 
\end{array}
\right), 
\end{equation}
and $\theta_{i} \in (0,\frac{2}{\lambda_{max}(A_{i}A_{i}^{T})}), i=1,2,...,m$.

\textit{Proof:} Rows of (\ref{2}) are written as $A_{i}x_{i}=b_{i}, i=1,2,...,m,$ which is equivalent to $x_{i}=x_{i}-\theta_{i} A_{i}^{T}(A_{i}x_{i}-b_{i})$. Consequently, the solution sets of $A_{i}x_{i}=b_{i}$ and $x_{i}=x_{i}-\theta_{i} A_{i}^{T}(A_{i}x_{i}-b_{i})$ are the same. This completes the proof of Lemma 7.  

\textbf{Remark 8:} Since $\lambda_{max}(A_{i}A_{i}^{T}) \leq \Vert A_{i}A_{i}^{T} \Vert_{\infty}, i=1, \hdots, m$, one may select $\theta_{i}=\frac{2}{\kappa_{i}}$ where $\kappa_{i} \geq \Vert A_{i}A_{i}^{T} \Vert_{\infty}$.

Now Problem 1 with Assumption 3 reduces to the following problem.

\textbf{Problem 2:} Consider Problem 1 with Assumption 3. Let $H(x):=\tilde{A}x+\tilde{b}$, where $\tilde{A}$ and $\tilde{b}$ are defined in (\ref{5})-(\ref{6}), and let $T(\omega^{*},x)$ be defined in Problem 1. The problem is to find $x^{*}$ such that $x^{*} \in Fix(H) \cap FVP(T).$ 

\textbf{Remark 9:} From Assumption 3, $Fix(H) \cap FVP(T) \neq \emptyset$. 

Now let $(\Omega^{*}, \sigma)$ be a measurable space where $\Omega^{*}$ and $\sigma$ are defined in Section II.B. Consider a probability measure $\mu$ defined on the space $(\Omega,\mathcal{F})$ where
$$\Omega=\Omega^{*} \times \Omega^{*} \times \Omega^{*} \times . . . $$
$$\mathcal{F}=\sigma \times \sigma \times \sigma \times . . . $$
such that $(\Omega,\mathcal{F},\mu)$ forms a probability space. We denote a realization in this probability space by $\omega \in \Omega$.

The Krasnoselskii-Mann iteration \cite{krasnos}-\cite{mann} for finding a fixed point of a nonexpansive operator $\Gamma(x)$ is
\begin{equation}\label{krasmann}
x_{n+1}=(1-\alpha_{n})x_{n}+\alpha_{n} \Gamma(x_{n})
\end{equation}
where $\alpha_{n} \in [0,1].$ The Picard iteration, which is (\ref{krasmann}) where $\alpha=1$, may not converge to a fixed point of $\Gamma(x)$, e.g., $\Gamma(x)=\Lambda x$ where $\Lambda=\begin{pmatrix}
0 & 0 & 1\\
1 & 0 & 0\\
0 & 1 & 0\\
\end{pmatrix}$ is periodic and irreducible. Krasnoselskii \cite{krasnos} proved that Algorithm (\ref{krasmann}) where $\alpha_{n}=\frac{1}{2}$ converges to a fixed point of $\Gamma(x)$.

We show in Lemma 9 that $Fix(H) \cap FVP(T) = FVP(D)$ where $D(\omega^{*},x):=(1-\beta) T(\omega^{*},x)+\beta H(x), \beta \in (0,1)$. Also we show in the proof of Lemma 10 that $D(\omega^{*},x)$ is nonexpansive. Hence, the random Krasnoselskii-Mann iterative algorithm for solving Problem 2 reduces to the following algorithm:

\begin{equation}\label{7}
x_{n+1}=\frac{1}{2} x_{n}+\frac{1}{2}[(1-\beta)W(\omega^{*}_{n})x_{n}+\beta (\tilde{A} x_{n} + \tilde{b})]
\end{equation}
where $\beta \in (0,1)$. 

Now we impose the following assumption on random communication graphs.

\textbf{Assumption 4} \cite{alavianiTAC}: There exists a nonempty subset $K \subseteq \Omega^{*}$ such that $FVP(T)=\{\tilde{z} | \tilde{z} \in X, \tilde{z}=W(\bar{\omega})\tilde{z}, \forall \bar{\omega} \in K  \}$. Moreover, $\bar{\omega} \in K$ occurs infinitely often almost surely.

\textbf{Remark 10} \cite{alavianiTAC}: If the sequence $\{ \omega_{n}^{*} \}_{n=0}^{\infty}$ is mutually independent with $\sum_{n=0}^{\infty} Pr_{n}(\bar{\omega})=\infty$ where $Pr_{n}(\bar{\omega})$ is the probability of occuring $\bar{\omega}$ at time $n$, then according to Borel-Cantelli lemma, Assumption 4 is satisfied. Moreover, any  ergodic stationary sequences $\{ \omega_{n}^{*} \}_{n=0}^{\infty}, Pr(\bar{\omega})>0,$ satisfy Assumption 4 (see proof of Lemma 1 in \cite{bikho}). Consequently, any time-invariant Markov chain with its unique stationary distribution as the initial distribution satisfy Assumption 4 (see \cite{bikho}).

Now we give our main theorem in this paper.

\textbf{Theorem 1:} Consider Problem 2 with Assumption 4. Then starting from any initial condition, the sequence generated by (\ref{7}) converges almost surely to $x^{*}$ which is the unique solution of the following convex optimization problem:
\begin{equation}\label{mainoptim}
\begin{aligned}
& \underset{x}{\text{min}}
& & \Vert x-x_{0} \Vert \\
& \text{subject to}
& & x=(1-\beta) W(\omega^{*}) x+\beta (\tilde{A} x+\tilde{b}), \; \forall \omega^{*} \in \Omega^{*}.
\end{aligned}
\end{equation}

\textbf{Remark 11:} Algorithm (\ref{7}) cannot be derived from generalization of algorithms proposed in \cite{lae1}-\cite{lae14} and \cite{elia}-\cite{cano} to random case.

Before we give the proof of Theorem 1, we should give some lemmas needed in the proof.

\textbf{Lemma 8:} Let $H(x)$ be defined in Problem 2. Then $H:\Re^{mq}  \longrightarrow \Re^{mq}$ is nonexpansive.

\textit{Proof:} See Appendix A.

\textbf{Lemma 9:} Let $T(\omega^{*},x)$ and $H(x)$ be defined in Problems 1 and 2, respectively, and 
\begin{equation}\label{operatord}
D(\omega^{*},x):=(1-\beta) T(\omega^{*},x)+\beta H(x), 
\end{equation}
where $\omega^{*} \in \Omega^{*}, \beta \in (0,1)$. Then $FVP(D)=Fix(H) \cap FVP(T)$.

\textit{Proof:} See Appendix B.

\textbf{Lemma 10:} Let $D(\omega^{*},x), \omega^{*} \in \Omega^{*}$, be defined in Lemma 9. Then $FVP(D)$ is a closed convex nonempty set.

\textit{Proof:} See Appendix C.

\textbf{Lemma 11:} Let $T(\omega^{*},x), \omega^{*} \in \Omega^{*}$, be defined in Problem 1, and 

\begin{equation}\label{lemma77}
S(\omega,x):=(1-\beta) T(\omega^{*},x)+\beta \tilde{A} x, \omega^{*} \in \Omega^{*},
\end{equation}
where $\beta \in (0,1)$. Then $FVP(S)$ is nonempty, closed, and convex.

\textit{Proof:} See Appendix D.

\textbf{Lemma 12:} Assume that the linear algebraic equation $Ax=b$ does not have the unique solution, i.e., $\mathcal{S}$ is not a singleton. Let $S(\omega^{*},x)$ be defined in (\ref{lemma77}). Then $FVP(S)$ is a closed affine subspace.

\textit{Proof:} See Appendix E.

\textbf{Lemma 13:} Let 
\begin{equation}\label{lmmla1}
Q_{1}(\omega^{*},x):=\frac{1}{2} x +\frac{1}{2} D(\omega^{*},x), \forall   \omega^{*} \in \Omega^{*},
\end{equation}
\begin{equation}\label{lmmla2}
Q_{2}(\omega^{*},x):=\frac{1}{2} x +\frac{1}{2} S(\omega^{*},x), \forall   \omega^{*} \in \Omega^{*}.
\end{equation}
Then $Q_{1}(\omega^{*},x)$ and $Q_{2}(\omega^{*},x)$ are nonexpansive and $FVP(Q_{1})=FVP(D)$ and $FVP(Q_{2})=FVP(S)$. Moreover, $Q_{1}(\omega^{*},x)$ is firmly nonexpansive for each $\omega^{*} \in \Omega^{*}$.  

\textit{Proof:} See Appendix F.

\textbf{Remark 12:} By Lemma 9 and Lemma 13, Assumption 3 guarantees that the set of equilibrium points of (\ref{7}) is $Fix(H) \cap FVP(T) \neq \emptyset.$ Also Assumption 3 guarantees the feasibility of the optimization problem (\ref{mainoptim}). 

\textbf{Remark 13:} Quadratic Lyapunov functions have been useful to analyze stability of linear dynamical systems. Nevertheless, quadratic Lyapunov functions may not exist for stability analysis of consensus problems in networked systems \cite{olshvtis}. Furthermore, quadratic Lyapunov functions may not exist for stability analysis of switched linear systems \cite{switch1}-\cite{switch3}. Moreover, other difficulties mentioned in \cite{alavfranklin} may arise in using Lyapunov's direct method to analyze stability of dynamical systems. Furthermore, LaSalle-type theorem for discrete-time stochastic systems (see \cite{lassle} and references therein) needs $\{ \omega^{*}_{n} \}_{n=0}^{\infty}$ to be independent. Therefore, we do not try Lyapunov's and LaSalle's approaches to analyze the stability of the dynamical system (\ref{7}) in this paper.

\textit{Proof of Theorem 1:} 

From Lemmas 9 and 13, we can write (\ref{7}) as
\begin{equation}\label{zzaazz}
x_{n+1}=Q_{1}(\omega^{*}_{n},x_{n}).
\end{equation}
Consider a $\bar{c} \in FVP(D)=FVP(Q_{1})$. From Lemma 13, we have $\bar{c}=Q_{1}(\omega^{*},\bar{c})$. Hence, for all $\omega \in \Omega$, we have
$$\Vert x_{n+1}-\bar{c} \Vert=\Vert Q_{1}(\omega^{*}_{n},x_{n}) - Q_{1}(\omega^{*}_{n},\bar{c}) \Vert \leq \Vert x_{n}-\bar{c} \Vert,$$
which implies that the sequence $\{x_{n} \}$ is Fej\'{e}r monotone with respect to $FVP(D)$ (see Definition 6 and Lemma 10). Therefore, the sequence is bounded by Lemma 2 for all $\omega \in \Omega$. Since $m \in N$, $\bar{N}$ is finite, we obtain from (\ref{zzaazz}), Lemma 4, and Assumption 4 that $\{ x_{n} \}_{n=0}^{\infty}$ converges almost surely to a random variable supported by $FVP(Q_{1})=FVP(D)$ for any initial condition.

It remains to prove that $\{ x_{n} \}_{n=0}^{\infty}$ converges almost surely to the unique solution $x^{*}$. If Problem 2 has a unique solution, then $x^{*}$ is the only feasible point of the optimization (\ref{mainoptim}); otherwise, $FVP(S)$ is a closed affine subspace by Lemma 12. Consider a fixed $\tilde{y} \in FVP(D)=FVP(Q_{1})$. Thus $\tilde{y}=\frac{1}{2} \tilde{y}+\frac{1}{2} D(\omega^{*},\tilde{y})$ and $D(\omega^{*},\tilde{y})=\tilde{y}, \forall \omega^{*} \in \Omega^{*}$. We obtain from these facts and (\ref{7}) that
\begin{align}
x_{n+1}-\tilde{y} &=\frac{1}{2} (x_{n}-\tilde{y})+\frac{1}{2} (D(\omega^{*}_{n},x_{n})-\tilde{y}) \nonumber \\
&=\frac{1}{2} (x_{n}-\tilde{y})+\frac{1}{2} (D(\omega^{*}_{n},x_{n})-D(\omega^{*}_{n},\tilde{y})) \nonumber \\
&=\frac{1}{2} (x_{n}-\tilde{y})+\frac{1}{2} (S(\omega^{*}_{n},x_{n})-S(\omega^{*}_{n},\tilde{y})) \nonumber \\
&=\frac{1}{2} (x_{n}-\tilde{y})+\frac{1}{2} S(\omega^{*}_{n},x_{n}-\tilde{y}) \nonumber \\
& =Q_{2}(\omega^{*}_{n},x_{n}-\tilde{y}) \label{sddsx}.
\end{align}

Now consider a $\bar{c} \in FVP(S)=FVP(Q_{2})$. From (\ref{sddsx}) we obtain
\begin{align*}
\Vert x_{n+1}-\tilde{y}-\bar{c} \Vert &=\Vert Q_{2}(\omega^{*}_{n},x_{n}-\tilde{y})-\bar{c} \Vert \nonumber \\
&=\Vert Q_{2}(\omega^{*}_{n},x_{n}-\tilde{y}) -Q_{2}(\omega^{*}_{n},\bar{c}) \Vert 
\end{align*}
which by nonexpansivity property of $Q_{2}(\omega^{*},x)$ (see Lemma 13) implies
$$\Vert x_{n+1}-\tilde{y}-\bar{c} \Vert =\Vert Q_{2}(\omega^{*}_{n},x_{n}-\tilde{y})-Q_{2}(\omega^{*}_{n},\bar{c}) \Vert$$
\begin{equation}\label{ddssd}
\leq \Vert x_{n}-\tilde{y}-\bar{c} \Vert.
\end{equation}
Since $FVP(S)=FVP(Q_{2})$ (by Lemma 13) is nonempty, closed, and convex (see Lemma 11), the sequence $\{ x_{n}-\tilde{y} \}_{n=0}^{\infty}$ is Fej\'{e}r monotone with respect to $FVP(Q_{2})=FVP(S)$ for all $\omega \in \Omega$. Moreover, $FVP(S)=FVP(Q_{2})$ (by Lemma 13) is a closed affine subspace by Lemma 12. Therefore, according to Lemma 3, we obtain
$$\lim_{n \longrightarrow \infty} x_{n}-\tilde{y}=P_{FVP(S)}(x_{0}-\tilde{y}).$$
As a matter of fact, $x^{*}=z^{*}+\tilde{y}$ where $z^{*}=P_{FVP(S)}(x_{0}-\tilde{y})$. Indeed, $z^{*}$ can be considered as the solution of the following convex optimization problem:  
\begin{equation}\label{optim1}
\begin{aligned}
& \underset{z}{\text{min}}
& & \Vert z-(x_{0}-\tilde{y}) \Vert \\
& \text{subject to}
& & z=(1-\beta) W(\omega^{*}) z+\beta \tilde{A} z, \; \forall \omega^{*} \in \Omega^{*}.
\end{aligned}
\end{equation}
By changing variable $x=z+\tilde{y}$ in optimization problem (\ref{optim1}), (\ref{optim1}) becomes
\begin{equation}\label{optim2}
\begin{aligned}
& \underset{x}{\text{min}}
& & \Vert x-x_{0} \Vert \\
& \text{subject to}
& & x=(1-\beta) W(\omega^{*}) (x-\tilde{y})+\beta \tilde{A} (x-\tilde{y})+\tilde{y}, \\
&&&   \forall \omega^{*} \in \Omega^{*}. 
\end{aligned}
\end{equation}
\begin{figure}[thpb]
	\centering
	\includegraphics[scale=0.5]{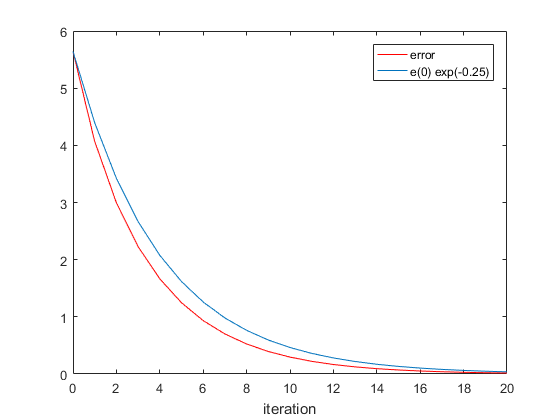}
	\caption{error}
	\label{error1example3}
\end{figure}
where $x^{*}$ is the solution of (\ref{optim2}). By the fact that $\tilde{y}=(1-\beta) \tilde{y}+\beta \tilde{y}$, the constraint set in (\ref{optim2}) becomes
\begin{equation}\label{ff1}
x=(1-\beta) (W(\omega^{*}) (x-\tilde{y})+\tilde{y})+\beta ( \tilde{A} (x-\tilde{y})+\tilde{y}),  \forall \omega^{*} \in \Omega^{*}. 
\end{equation}
Substituting $\tilde{y}=W(\omega^{*}) \tilde{y}, \forall \omega^{*} \in \Omega^{*}$, and $\tilde{y}=\tilde{A} \tilde{y}+\tilde{b}$ for (\ref{ff1}) yields
\begin{equation}\label{fffif1}
x=(1-\beta) W(\omega^{*}) x +\beta (\tilde{A} x+\tilde{b}).
\end{equation}
Substituting (\ref{fffif1}) for (\ref{optim2}) yields (\ref{mainoptim}). Because of strict convexity of 2-norm $\Vert . \Vert$, the convex optimization problem (\ref{mainoptim}) has the unique solution. Thus the proof of Theorem 1 is complete.

\textbf{Remark 14:} The definition of fixed value point is a bridge from deterministic analysis to random analysis of the algorithm (see \cite{alavianiTAC}). With the help of fixed value point set and nonexpansivity property of the random operator $D(\omega^{*},x)$, we are able to prove boundedness of the generated sequence in a deterministic way and to use deterministic tools such as Lemmas 3 and 4 to prove the convergence of the algorithm in the proof of Theorem 1. Therefore, the definition of fixed value point set with nonexpansivity property of $D(\omega^{*},x)$ makes analysis of random processes easier than those of existing results regardless of switching distributions. This is very useful because we are able to analyze random processes by using extended deterministic tools (see also \cite{alavianiTAC} for optimization problems).

\textbf{Theorem 2:} Consider Problem 2 with Assumption 4. Then starting from any initial condition, the sequence generated by (\ref{7}) converges in mean square to $x^{*}$ which is the unique solution of the convex optimization problem (\ref{mainoptim}).

\textit{Proof:} See Appendix G.

\textbf{Remark 15:} The solution of the optimization problem (\ref{mainoptim}) is independent of the choice $\beta$.

\textbf{Remark 16:} The optimization problem (\ref{mainoptim}) is equivalent to the following optimization problem:

\begin{equation}\label{optend}
\begin{aligned}
& \underset{x}{\text{min}}
& & \Vert x-x_{0} \Vert \\
& \text{subject to} & &  x=W(\omega^{*}) x, \; \forall \omega^{*} \in \Omega^{*}, \\
& & &  x=\tilde{A} x+\tilde{b}. \\
\end{aligned}
\end{equation}

\begin{figure}[thpb]
	\centering
	\includegraphics[scale=0.5]{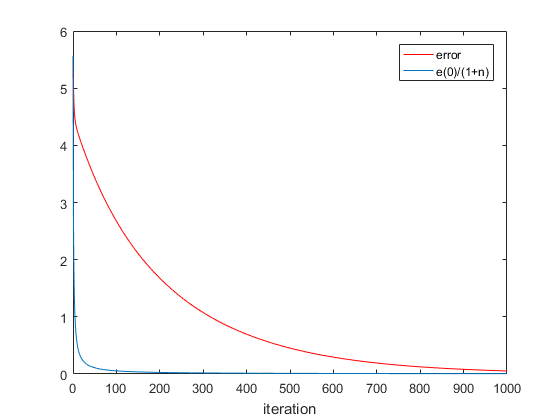}
	\caption{error}
	\label{error2example3}
\end{figure}

\textbf{Remark 17:} Since Assumption 2 holds, the constraint set $\{x | x=W(\omega^{*})x, \forall \omega^{*} \in \Omega^{*}  \}$ in (\ref{optend}) does not depend on choices of links' weights. This fact implies that the limit point $x^{*}$ of Algorithm (\ref{7}) is not determined by any choices of links' weights as long as Assumptions 2-4 are satisfied; consequently, the limit point $x^{*}$ is robust to any uncertainties of links' weights.

\textbf{Remark 18:} The rate of convergence of Algorithm (\ref{7}) cannot be guaranteed (see Examples 1 and 2).

\section{Numerical Examples}

\textit{Example 1:} Consider three agents which want to solve a linear algebraic equation \[A\left(
\begin{array}{c c c}
x\\
y\\
z\\
\end{array}
\right)=b, A=\left(
\begin{array}{c c c}
1 & 2 & 1\\
2 & 4 & 2\\
3 & 6 & 3\\
\end{array}
\right), b=\left(
\begin{array}{c c c}
1\\
2\\
3\\
\end{array}
\right).\]
Clearly, $\{[x,y,z]^{T} \in \Re^{3} | x+2y+z=1  \}$ is the solution set. Each agent $i, i=1,2,3$, only knows the $i$-th row of $[A,b]$. We consider undirected link between any two agents, namely a complete graph, where the weight of each link is $\frac{1}{3}$. We assume that the communication graph is non-random. Thus the conditions of Theorem 1 are fulfilled. We choose $\beta=0.5,\theta_{1}=\frac{1}{6}, \theta_{2}=\frac{1}{24}, \theta_{3}=\frac{1}{54}$, and initial conditions $x_{1}(0)=-3,y_{1}(0)=1,z_{1}(0)=2,x_{2}(0)=2,y_{2}(0)=-2,z_{2}(0)=1,x_{3}(0)=1,y_{3}(0)=3, z_{3}(0)=-1$ for simulation. We use CVX software of Matlab to solve the optimization (\ref{mainoptim}), and the solution is
$x^{*}=\textbf{1}_{3} \otimes \begin{pmatrix}
-0.1667\\
0.3333\\
0.5000\\
\end{pmatrix}.$
Then the error $e_{n}=\Vert x_{n}-x^{*} \Vert$ converges to zero exponentially fast with decay rate $0.25$ where the result is shown in Figure (\ref{error1example3}).

\textit{Example 2:} Consider three agents which want to solve a linear algebraic equation

\[A\left(
\begin{array}{c c c}
x\\
y\\
z\\
\end{array}
\right)=b, A=\left(
\begin{array}{c c c}
1 & 0 & 0\\
2 & 1 & 0\\
3 & 1 & 2\\
\end{array}
\right), b=\left(
\begin{array}{c c c}
1\\
2\\
1\\
\end{array}
\right).\]
Clearly, $[1,0,-1]^{T}$ is the unique solution. Each agent $i, i=1,2,3$, only knows the $i$-th row of $[A,b]$. We consider undirected link between any two agents, namely a complete graph, where the weight of each link is $\frac{1}{3}$. We select $\beta=0.5,\theta_{1}=1, \theta_{2}=\frac{1}{5},$ and $\theta_{3}=\frac{1}{14}$, and random initial conditions for simulation.

Then the error $e_{n}=\Vert x_{n}-x^{*} \Vert$ converges to zero slower than $\frac{e(0)}{1+n}$ where the result is shown in Figure (\ref{error2example3}).

Note that Examples 1 and 2 show that the rate of convergence of Algorithm (\ref{7}) cannot be guaranteed.

\section{Conclusion}

In this paper, we consider the problem of solving linear algebraic equations of the form $Ax=b$ over a network of multi-agent systems. The equation is solved by $m$ agents where each agent only knows a subset of rows of the partitioned matrix $[A,b]$ in presence of random communication topologies. We formulate the problem in a way that the distribution of random communication graphs or B-connectivity assumption is not needed. Hence, this formulation includes asynchronous updates or unreliable communication protocols. We apply the random Krasnoselskii-Mann iteration which converges almost surely and in mean square to a solution of the problem for any matrices $A$ and $b$ and any intial conditions of agents' states if a solution exists. We show that the limit point to which all agents' states converge is determined by the unique solution of a convex optimization problem. Ultimately, two numerical examples are given to validate that the rate of convergence of the algorithm cannot be guaranteed.


%
\appendices

\section{}

\textit{Proof of Lemma 8:} We have that $\Vert H(z)-H(y) \Vert = \Vert \tilde{A} (z-y) \Vert, \forall z,y \in \Re^{mq}$. Now we prove that $\Vert \tilde{A}(z-y) \Vert \leq \Vert z-y \Vert.$ Let $z=[z_{1},z_{2},...,z_{m}]^{T}$ and $y=[y_{1},y_{2},...,y_{m}]^{T}$. We have that
$$\Vert \tilde{A}(z-y) \Vert^{2}$$
\[= \Vert \left(
\begin{array}{c c c c}
(I_{q}-\theta_{1} A_{1}^{T}A_{1})(z_{1}-y_{1})\\
(I_{q}-\theta_{2} A_{2}^{T}A_{2})(z_{2}-y_{2})\\
\vdots\\
(I_{q}-\theta_{m} A_{m}^{T}A_{m})(z_{m}-y_{m})\\ 
\end{array}
\right) \Vert^{2}\] 
$$=\sum_{j=1}^{m} \Vert (I_{q}-\theta_{j} A_{j}^{T}A_{j})(z_{j}-y_{j}) \Vert^{2}.$$
Since $\theta_{j} \in (0,\frac{2}{\lambda_{max}(A_{j}A_{j}^{T})})$, we have $\Vert I_{q}-\theta_{j} A_{j}^{T}A_{j} \Vert \leq 1$. Moreover, $\Vert (I_{q}-\theta_{j} A_{j}^{T}A_{j})(z_{j}-y_{j}) \Vert \leq \Vert I_{q}-\theta_{j} A_{j}^{T}A_{j} \Vert \Vert z_{j}-y_{j} \Vert, j=1,2,...,m$. Therefore, we obtain
$$\sum_{j=1}^{m} \Vert (I_{q}-\theta_{j} A_{j}^{T}A_{j})(z_{j}-y_{j}) \Vert^{2}$$
$$\leq \sum_{j=1}^{m} \Vert I_{q}-\theta_{j} A_{j}^{T}A_{j} \Vert^{2} \Vert z_{j}-y_{j} \Vert^{2} $$
$$\leq \sum_{j=1}^{m}  \Vert z_{j}-y_{j} \Vert^{2}=\Vert z-y \Vert^{2}$$
or
\begin{equation}\label{11}
\Vert \tilde{A}(z-y) \Vert \leq \Vert z-y \Vert.
\end{equation}
Thus the proof of Lemma 8 is complete.

\section{}

\textit{Proof of Lemma 9:} Assume a $\tilde{z} \in Fix(H) \cap FVP(T)$. In fact, $\tilde{z}=H(\tilde{z})$ and $\tilde{z}=T(\omega^{*},\tilde{z})=\tilde{z}, \forall \omega^{*} \in \Omega^{*}$. Therefore, we obtain from (\ref{operatord}) that
\begin{align*}
D(\omega^{*},\tilde{z}) &=(1-\beta) T(\omega^{*},\tilde{z})+\beta H(\tilde{z}) \nonumber \\
&=(1-\beta)\tilde{z}+\beta \tilde{z}=\tilde{z}, \forall \omega^{*} \in \Omega^{*}, 
\end{align*}
which implies that $Fix(H) \cap FVP(T) \subseteq FVP(D)$. Conversely, assume a $\tilde{z} \in FVP(D)$, i.e.,
\begin{equation}\label{dzz}
D(\omega^{*},\tilde{z})=\tilde{z}=(1-\beta) T(\omega^{*},\tilde{z})+\beta H(\tilde{z}), \forall \omega^{*} \in \Omega^{*}.
\end{equation}
Since $Fix(H) \cap FVP(T) \neq \emptyset$, there exits a $y^{*} \in Fix(H) \cap FVP(T)$. Now by (\ref{dzz}) we obtain
$$\Vert \tilde{z}-y^{*} \Vert=\Vert (1-\beta) T(\omega^{*},\tilde{z})+\beta H(\tilde{z})- y^{*} \Vert.$$
By the fact that $y^{*}=(1-\beta)y^{*}+\beta y^{*}, \beta \in (0,1)$, we obtain
\begin{align}
\Vert \tilde{z}-y^{*} \Vert &=\Vert (1-\beta) T(\omega^{*},\tilde{z})+\beta H(\tilde{z})- y^{*} \Vert \nonumber \\
&=\Vert (1-\beta) (T(\omega^{*},\tilde{z})-y^{*})+\beta (H(\tilde{z})-y^{*}) \Vert. \label{ppp1}
\end{align}
Since $y^{*}=H(y^{*})$ and $y^{*}=T(\omega^{*},y^{*}), \forall \omega^{*} \in \Omega^{*}$, we obtain from (\ref{ppp1}) for all $\omega^{*} \in \Omega^{*}$ that
$$\Vert (1-\beta) (T(\omega^{*},\tilde{z})-y^{*})+\beta (H(\tilde{z})-y^{*}) \Vert=$$
\begin{equation}\label{sttt1}
\Vert (1-\beta) (T(\omega^{*},\tilde{z})-T(\omega^{*},y^{*}))+\beta (H(\tilde{z})-H(y^{*})) \Vert. 
\end{equation}
Due to nonexpansivity property of $T(\omega^{*},x)$, we have that
$$\Vert (1-\beta) (T(\omega^{*},\tilde{z})-T(\omega^{*},y^{*}))+\beta (H(\tilde{z})-H(y^{*})) \Vert$$ 
\begin{equation}\label{sttt2}
\leq (1-\beta) \Vert \tilde{z}-y^{*} \Vert + \beta \Vert H(\tilde{z})-H(y^{*}) \Vert, \forall \omega^{*} \in \Omega^{*}.
\end{equation}
By nonexpansivity property of $H(x)$ (see Lemma 8), we also have that
$$\Vert (1-\beta) (T(\omega^{*},\tilde{z})-T(\omega^{*},y^{*}))+\beta (H(\tilde{z})-H(y^{*})) \Vert$$ 
\begin{equation}\label{sttt3}
\leq (1-\beta) \Vert T(\omega^{*},\tilde{z})-T(\omega^{*},y^{*}) \Vert + \beta \Vert \tilde{z}-y^{*} \Vert, \forall \omega^{*} \in \Omega^{*}.
\end{equation}
Because of nonexpansivity property of $T(\omega^{*},x)$, we obtain from (\ref{sttt3}) that
$$(1-\beta) \Vert T(\omega^{*},\tilde{z})-T(\omega^{*},y^{*}) \Vert + \beta \Vert \tilde{z}-y^{*} \Vert$$ 
\begin{equation}\label{sttt4}
\leq (1-\beta) \Vert \tilde{z}-y^{*} \Vert + \beta \Vert \tilde{z}-y^{*} \Vert=\Vert \tilde{z}-y^{*} \Vert, \forall \omega^{*} \in \Omega^{*}. 
\end{equation}
Due to nonexpansivity property of $H(x)$, we also obtain from (\ref{sttt2}) that
$$(1-\beta) \Vert \tilde{z}-y^{*} \Vert + \beta \Vert H(\tilde{z})-H(y^{*}) \Vert$$
\begin{equation}\label{sttt5}
\leq (1-\beta) \Vert \tilde{z}-y^{*} \Vert + \beta \Vert \tilde{z}-y^{*} \Vert=\Vert \tilde{z}-y^{*} \Vert. 
\end{equation}
From (\ref{ppp1})-(\ref{sttt5}), we finally obtain
$$\Vert \tilde{z}-y^{*} \Vert \leq$$
$$\Vert (1-\beta) (T(\omega^{*},\tilde{z})-T(\omega^{*},y^{*}))+\beta (H(\tilde{z})-H(y^{*})) \Vert$$
$$\leq (1-\beta) \Vert \tilde{z} - y^{*} \Vert+ \beta \Vert H(\tilde{z})-H(y^{*}) \Vert$$
\begin{equation}\label{ppll1}
\leq \Vert \tilde{z}-y^{*} \Vert, \forall \omega^{*} \in \Omega^{*}
\end{equation}
and 
$$\Vert \tilde{z}-y^{*} \Vert \leq$$
$$\Vert (1-\beta) (T(\omega^{*},\tilde{z})-T(\omega^{*},y^{*}))+\beta (H(\tilde{z})-H(y^{*})) \Vert$$
$$\leq (1-\beta) \Vert T(\omega^{*}, \tilde{z})-T(\omega^{*}, y^{*}) \Vert + \beta \Vert \tilde{z}-y^{*} \Vert$$
\begin{equation}\label{ppll2}
\leq \Vert \tilde{z}-y^{*} \Vert, \forall \omega^{*} \in \Omega^{*}.
\end{equation}
Thus, the equalities hold in (\ref{ppll1}) and (\ref{ppll2}), that imply that
$$\Vert \tilde{z}-y^{*} \Vert$$
$$=\Vert (1-\beta) (T(\omega^{*},\tilde{z})-T(\omega^{*},y^{*}))+\beta (H(\tilde{z})-H(y^{*})) \Vert$$
$$=\Vert H(\tilde{z})-H(y^{*}) \Vert$$
\begin{equation}\label{ppll3}
= \Vert T(\omega^{*},\tilde{z})-T(\omega^{*},y^{*}) \Vert, \forall \omega^{*} \in \Omega^{*}.
\end{equation}
Substituting $y^{*}=H(y^{*})$ and $y^{*}=T(\omega^{*},y^{*}), \forall \omega^{*} \in \Omega^{*}$, for (\ref{ppll3}) yields
$$\Vert H(\tilde{z})-y^{*} \Vert= \Vert T(\omega^{*},\tilde{z})-y^{*} \Vert=$$
$$\Vert (1-\beta) (T(\omega^{*},\tilde{z})-y^{*})+\beta (H(\tilde{z})-y^{*}) \Vert,\forall \omega^{*} \in \Omega^{*},$$
which by Remark 2 implies that $H(\tilde{z}) - y^{*}=T(\omega^{*},\tilde{z})-y^{*}, \forall \omega^{*} \in \Omega^{*},$ or 
\begin{equation}\label{fiiix}
H(\tilde{z})=T(\omega^{*},\tilde{z}),\forall \omega^{*} \in \Omega^{*}.
\end{equation}
Substituting (\ref{fiiix}) for (\ref{dzz}) yields
$$\tilde{z}=H(\tilde{z})=T(\omega^{*},\tilde{z}), \forall \omega^{*} \in \Omega^{*},$$
which implies that $FVP(D) \subseteq Fix(H) \cap FVP(T)$. Therefore, $FVP(D)=Fix(H) \cap FVP(T)$. Thus the proof of Lemma 9 is complete.

\section{}

\textit{Proof of Lemma 10:} For any $z,y \in \Re^{mq}$, we obtain
$$\Vert D(\omega^{*},z)-D(\omega^{*},y) \Vert$$
$$=\Vert (1-\beta)(T(\omega^{*},z)-T(\omega^{*},y))+\beta (H(z)-H(y)) \Vert$$
\begin{equation}\label{ddfvp1}
\leq (1-\beta) \Vert T(\omega^{*},z)-T(\omega^{*},y) \Vert+\beta \Vert H(z)-H(y) \Vert.
\end{equation}
Because of nonexpansivity of both $T(\omega^{*},x)$ and $H(x)$, we obtain from (\ref{ddfvp1}) that
$$\Vert D(\omega^{*},z)-D(\omega^{*},y) \Vert $$
$$\leq (1-\beta) \Vert T(\omega^{*},z)-T(\omega^{*},y) \Vert+\beta \Vert H(z)-H(y) \Vert$$
$$\leq (1-\beta) \Vert z-y \Vert + \beta \Vert z-y \Vert= \Vert z-y \Vert$$
that implies that $D(\omega^{*},x)$ is nonexpansive. Indeed, since $\Re^{mq}$ is closed (see Preposition 1) and convex, we obtain by Remark 1 that $FVP(D)$ is closed and convex. Furthermore, $FVP(D)$ is nonempty by Assumption 3 and Lemma 9. This completes the proof of Lemma 10.

\section{}

\textit{Proof of Lemma 11:} Since $\textbf{0}_{mq}$ is a fixed value point of $S$, we can conclude that $FVP(S)$ is nonempty. Now for any $z,y \in \Re^{mq}$, we obtain
$$\Vert S(\omega^{*},z)-S(\omega^{*},y) \Vert$$
$$=\Vert (1-\beta)(T(\omega^{*},z)-T(\omega^{*},y))+\beta \tilde{A}(z-y) \Vert$$
\begin{equation}\label{lemmm1}
\leq (1-\beta) \Vert T(\omega^{*},z)-T(\omega^{*},y) \Vert +\beta \Vert \tilde{A}(z-y) \Vert.
\end{equation}
Similar to the proof of Lemma 8, we obtain
\begin{equation}\label{lemmm2}
\Vert \tilde{A}(z-y) \Vert \leq \Vert z-y \Vert.
\end{equation}
Therefore, we obtain from (\ref{lemmm1}) by nonexpansivity of $T(\omega^{*},x)$ and (\ref{lemmm2}) that
$$\Vert S(\omega^{*},z)-S(\omega^{*},y) \Vert$$
$$\leq (1-\beta) \Vert T(\omega^{*},z)-T(\omega^{*},y) \Vert +\beta \Vert \tilde{A}(z-y) \Vert$$
$$\leq (1-\beta) \Vert z-y \Vert+\beta \Vert z-y \Vert=\Vert z-y \Vert$$
which implies that $S(\omega^{*},x),\omega^{*} \in \Omega^{*}$, is nonexpansive. Therefore, one can obtain by Remark 1 that $FVP(S)$ is closed and convex. Thus the proof of Lemma 11 is complete.

\section{}

\textit{Proof of Lemma 12:} By Lemma 11, we have that $FVP(S)$ is closed. Since $\mathcal{S}$ is not a singleton, $FVP(S)$ is not a singleton either. Consider two distinct points $\bar{z},\bar{y} \in FVP(S)$, i.e., 
\begin{equation}\label{nmmm1}
\bar{z}=S(\omega^{*},\bar{z}),\bar{y}=S(\omega^{*},\bar{y}), \forall \omega^{*} \in \Omega^{*}.
\end{equation}
Now we obtain
\begin{align}
S(\omega^{*},\alpha \bar{z}+(1-\alpha) \bar{y}) &=S(\omega^{*},\alpha \bar{z})+S(\omega^{*},(1-\alpha) \bar{y}) \nonumber \\
&=\alpha S(\omega^{*},\bar{z})+(1-\alpha) S(\omega^{*},\bar{y}), \label{nmmm2}
\end{align}
where $\alpha \in \Re$. Substituting (\ref{nmmm1}) for (\ref{nmmm2}) yields
$$S(\omega^{*},\alpha \bar{z}+(1-\alpha) \bar{y})=$$
$$\alpha S(\omega^{*},\bar{z})+(1-\alpha) S(\omega^{*},\bar{y})=\alpha \bar{z}+(1-\alpha) \bar{y}$$
which implies that $\alpha \bar{z}+(1-\alpha) \bar{y} \in FVP(S)$. Therefore, $FVP(S)$ is an affine set. Since $\textbf{0}_{mq} \in FVP(S)$, we obtain  by Remark 3 that the set
$$FVP(S)-\textbf{0}_{mq}=FVP(S)$$
is a subspace. Thus the proof of Lemma 12 is complete.

\section{}

\textit{Proof of Lemma 13:} Since $D(\omega^{*},x)$ and $S(\omega^{*},x)$ are nonexpansive, we obtain by Remark 4 that $Q_{1}(\omega^{*},x)$ and $Q_{1}(\omega^{*},x)$ are firmly nonexpansive for each $\omega^{*} \in \Omega^{*}$ and thus nonexpansive. Now consider a $\tilde{z} \in FVP(D)$. Thus $D(\omega^{*},\tilde{z})=\tilde{z}, \forall \omega^{*} \in \Omega^{*}$. Substituting this fact for (\ref{lmmla1}) yields $Q_{1}(\omega^{*},\tilde{z})=\tilde{z}, \forall \omega^{*} \in \Omega^{*}$ which implies that $\tilde{z} \in FVP(Q_{1})$. Now consider a $\tilde{z} \in FVP(Q_{1})$. Similarly, one can obtain that $\tilde{z} \in FVP(D)$. Therefore, $FVP(Q_{1})=FVP(D)$. With the same procedure, one can prove by using nonexpansivity of $S(\omega,x)$ (see proof of Lemma 11) that $FVP(Q_{2})=FVP(S)$. Thus the proof of Lemma 13 is complete.

\section{}

\textit{Proof of Theorem 2:} We have from Theorem 1 that $\displaystyle \lim_{n \longrightarrow \infty} \Vert x_{n}-x^{*} \Vert=0$ almost surely, or $\displaystyle \lim_{n \longrightarrow \infty} \Vert x_{n}-x^{*} \Vert^{2}=0$ almost surely. From Parallelogram Law, we have that
$$\Vert x_{n}-x^{*} \Vert^{2} \leq 2 (\Vert x_{n} \Vert^{2}+\Vert x^{*} \Vert^{2}), \forall n \in N.$$ 
We define a nonnegative measurable function $\tau_{n}=2 (\Vert x_{n} \Vert^{2}+\Vert x^{*} \Vert^{2})-\Vert x_{n}-x^{*} \Vert^{2}.$ Hence, $\displaystyle \lim_{n \longrightarrow \infty} \tau_{n}=4 \Vert x^{*} \Vert^{2}$ almost surely. Applying Lemma 5 yields
$$\int_{\Omega} (\liminf_{n \longrightarrow \infty} \tau_{n}) d \mu \leq \liminf_{n \longrightarrow \infty} \int_{\Omega} \tau_{n} d \mu$$
or
$$\int_{\Omega} 4 \Vert x^{*} \Vert^{2} d \mu \leq $$
\begin{equation}\label{dominated}
\liminf_{n \longrightarrow \infty} (\int_{\Omega} 2 \Vert x_{n} \Vert^{2} d \mu+ \int_{\Omega} 2 \Vert x^{*} \Vert^{2} d \mu-\int_{\Omega} \Vert x_{n}-x^{*} \Vert^{2} d \mu). 
\end{equation}
Due to boundedness of $\{ x_{n}  \}_{n=0}^{\infty}, \forall \omega \in \Omega,$ we obtain by Lemma 6 that $\displaystyle \lim_{n \longrightarrow \infty} \int_{\Omega} 2 \Vert x_{n} \Vert^{2} d \mu=\int_{\Omega} 2 \Vert x^{*} \Vert^{2} d \mu.$ Thus, we obtain from this fact and (\ref{dominated}) that 
$$\int_{\Omega} 4 \Vert x^{*} \Vert^{2} d \mu \leq$$
$$ \liminf_{n \longrightarrow \infty} (\int_{\Omega} 2 \Vert x_{n} \Vert^{2} d \mu+ \int_{\Omega} 2 \Vert x^{*} \Vert^{2} d \mu-\int_{\Omega} \Vert x_{n}-x^{*} \Vert^{2} d \mu)=$$
$$ \lim_{n \longrightarrow \infty} (\int_{\Omega} 2 \Vert x_{n} \Vert^{2} d \mu)+ \int_{\Omega} 2 \Vert x^{*} \Vert^{2} d \mu-\limsup_{n \longrightarrow \infty} \int_{\Omega} \Vert x_{n}-x^{*} \Vert^{2} d \mu $$
or
$$\limsup_{n \longrightarrow \infty} \int_{\Omega} \Vert x_{n}-x^{*} \Vert^{2} d \mu=0.$$
Therefore, we obtain
$$\lim_{n \longrightarrow \infty} E[\Vert x_{n}-x^{*} \Vert^{2}] \leq \limsup_{n \longrightarrow \infty} \int_{\Omega} \Vert x_{n}-x^{*} \Vert^{2} d \mu=0$$
which implies that $\{ x_{n}  \}_{n=0}^{\infty}$ converges in mean square to $x^{*}$. Thus the proof of Theorem 2 is complete.

%


\end{document}